\documentstyle[12pt,epsf]{article} 

\def\lsim{\mathrel{\lower2.5pt\vbox{\lineskip=0pt\baselineskip=0pt 
           \hbox{$<$}\hbox{$\sim$}}}} 
\def\gsim{\mathrel{\lower2.5pt\vbox{\lineskip=0pt\baselineskip=0pt 
           \hbox{$>$}\hbox{$\sim$}}}} 
\def\dis{\displaystyle}
\def\G{\Gamma}

\def\p{\partial}

\def\L{\Lambda}

\def\z{\zeta}
\def\a{\alpha}
\def\b{\beta}
\begin{document} 
\begin{flushright}
DPNU-02-42\\ hep-ph/0301168
\end{flushright}

\vspace{10mm}

\begin{center}
{\Large \bf 
 Casimir Energies due to Matter Fields in $T^{2}$ and $T^{2}/Z_{2}$
 Compactifications}

\vspace{20mm}
 Masato Ito 
 \footnote{E-mail address: mito@eken.phys.nagoya-u.ac.jp}
\end{center}

\begin{center}
{
\it 
{}Department of Physics, Nagoya University, Nagoya, 
JAPAN 464-8602 
}
\end{center}

\vspace{25mm}

\begin{abstract}
 \baselineskip=6mm
 We calculate the Casimir energies due to matter fields with various
 boundary conditions along two compact directions in $T^{2}$
 compactification.
 We discuss whether the Casimir energies generate attractive or
 repulsive forces.
 On the theories with extra dimensions,
 the Casimir energy plays a crucial role in the
 mechanism for stabilizing the size of extra dimensions.
 Finally we argue a procedure of the application to $Z_{2}$ orbifold.
\end{abstract} 

\newpage 
 \section{Introduction}
 \baselineskip=7mm

 Motivated by Kaluza \cite{Kaluza} and Klein \cite{Klein},
 it is expected that the phenomenology
 of low energy physics should be explained by extra spatial dimensions.
 Based on the Kaluza-Klein theory it is natural to consider that
 the radii of the compactified extra dimensions should be Planck size.
 After the Kaluza-Klein theory, there had been many works in the
 framework of theories with extra dimensions
 \cite{KK,Appelquist:1983vs,Antoniadis:1998ig,Antoniadis:1990ew}.
 Relying on the assumption that the gravity only feels large
 extra compact dimensions at sub-millimeter range, a remarkable model of
 Ref. \cite{Arkani-Hamed:1998rs} provided a breakthrough of hierarchy
 problem.

 Recent developments are based on the idea that ordinary matter
 fields could be confined to a three-brane world embedded in the
 higher dimensional world.
 The assumption of such models is that the three-brane can be identified
 with the fixed plane via orbifold of extra dimensional manifold.
 At the beginning of warped braneworld scenario, Randall and Sundrum had
 proposed a new suggestion to the hierarchy problem by using separated
 three-branes on $S^{1}/Z_{2}$ orbifold embedded in $AdS_{5}$
 \cite{Randall:1999ee}.
 Furthermore, it was shown that the localization of the gravity occurs on
 the positive tension three-brane and usual four-dimensional Newton law
 can be recovered at distance which is much larger than radius of 
 $AdS_{5}$ \cite{Randall:1999vf}.
 Thus the orbifold of compact spaces may be very fascinating idea in the
 framework of braneworld scenarios.  

 Interestingly, an operation of the orbifold is very useful to
 construct phenomenological models because unwanted matter fields can
 be projected out, and
 the operation enables for the models to be chiral. 
 In order that the model can be phenomenologically viable,
 it is possibility that wanted matter fields can be assigned to
 appropriate charges under orbifold symmetry.
 Furthermore it is widely discussed that the orbifold brings on the
 supersymmetry breaking or the breaking of gauge group
 \cite{Kawamura:1999nj,Barbieri:2000vh}.
 The idea of the orbifold had been developed by string theory 
 in nontrivial background.
 Recent phenomenological successful models via the orbifold are not 
 necessary based on string theory.
 However it is expected that these remarkable models can provide several
 solutions to triplet-doublet splitting, proton decay, neutrino models
 and so on.
 
 On the theories with extra dimensions, an important issue arises.
 The stabilization mechanism for the size of the extra spatial
 dimensions has not been discovered yet.
 As a clue to the issue, the Casimir energy plays a crucial
 role in radius stabilization of models with compactified extra
 dimensions \cite{Hofmann:2000cj,Ponton:2001hq}.
 The fate of compactified extra dimensions depends on the Casimir force
 generated by the Casimir energy. 
 Attractive force leads to shrink of the compactified manifold,
 on the other hand, repulsive force leads to inflation of the compactified
 manifold.
 Accordingly it is considered that radius stabilization can be realized
 by balance between the attractive and the repulsive force.

 In this paper, we calculate the Casimir energy $V$ for a massless
 scalar field in $M^{4}\times T^{2}$, where $M^{4}$ is ordinary
 four-dimensional Minkowski space and $T^{2}$ is two-torus.
 It is assumed that the fifth and the sixth dimensions are compactified.
 Note that the Casimir energy depends on the topology and boundary
 conditions of compactified manifold, and we consider that the boundary
 condition is periodic or anti-periodic along each compactified direction.
 We focus on the Casimir energies with respect to the following four
 cases, $V^{(P,P)}$, $V^{(P,A)}$, $V^{(A,P)}$ and $V^{(A,A)}$, where
 $P$ denotes periodic boundary condition and $A$ does anti-periodic
 boundary condition.
 In the round bracket of $V$,
 the former letter and the latter letter represents the boundary
 condition for the fifth direction and for the sixth direction,
 respectively.  
 The evaluation of $V^{(P,P)}$ has been performed in Ref.
 \cite{Ponton:2001hq}, accordingly, we will evaluate the remaining three
 cases.
 By examining the sign of the Casimir energy, we will investigate
 whether the Casimir force is attractive or repulsive.
 Finally we discuss an application to $Z_{2}$ orbifold.
 Since we are interested in the Casimir energies due to the matter fields,
 in the present paper we neglect the contributions of the various
 localized terms on the orbifold fixed planes in addition to a bulk
 cosmological constant.  
 We are going to provide the formulas of the Casimir energies due to the
 matters fields when considering the radius stabilization of a certain
 model.

 The Casimir energy due to a scalar field with the Kaluza-Klein modes
 can be given by
 \begin{eqnarray}
 V&=&\frac{1}{2}\sum_{m,n}\int\frac{d^{4}k}{(2\pi)^{4}}
   \log\left(k^{2}+{\cal M}^{2}_{m,n}\right)\nonumber\\
 &=&-\frac{1}{2}\sum_{m,n}\left.\frac{\p}{\p s}\right|_{s=0}
     \int\frac{d^{4}k}{(2\pi)^{4}}
     \left(k^{2}+{\cal M}^{2}_{m,n}\right)^{-s}\nonumber\\
 &=&-\frac{1}{32\pi^{2}}\left.\frac{\p}{\p s}\right|_{s=-2}
     \frac{1}{s(s+1)}\sum_{m,n}\left({\cal M}^{2}_{m,n}\right)^{-s}
 \label{eqn1}\,,
 \end{eqnarray}
 where $m,n\in {\bf Z}$.
 Note that ${\cal M}^{2}_{m,n}$ corresponds to the Kaluza-Klein spectrum
 which is determined by boundary conditions.
 The evaluation of the double sum in (\ref{eqn1}) is explicitly
 performed by using the techniques of $\zeta$-function regularization
 \cite{zeta1,zeta2,Elizalde:fg,Itzykson:xw,kubota}.
 When we calculate the Casimir energy of matter field with various
 boundary conditions on $T^{2}$, we use the above formula.
 Applying our results to a certain phenomenological models,
 we multiply (\ref{eqn1}) by the number of degrees of freedom of scalar
 field in the model, while for fermion we need to multiply by $-1$.  

 The paper is organized as follows.
 In section 2 we describe the detailed calculations of the Casimir energies
 due to a massless scalar field with boundary conditions along two
 directions of $T^{2}$.
 The Casimir energies can be written in terms of the area of the torus
 and the ratio of two radii, and we study the Casimir forces.
 Finally we shall briefly discuss the case of $Z_{2}$ orbifold.
 In section 3 we mention a summary with respect to the results obtained
 in the paper.
 In Appendix A we show the detailed evaluations of the double sums
 including in calculations of the Casimir energies.
 In Appendix B we derive the relation between $\z^{\prime}(-2n)$ and
 $\z(2n+1)$ by calculating the Casimir energy on $S^{1}$
 compactification.
 Because we encounter the first derivative of zeta function when evaluating 
 the Casimir energy.
 In final part, a procedure of the evaluation of $\z^{\prime}(0)$ is
 described.

%
 \section{Calculation of the Casimir energies}

 We consider $T^{2}$ compactification in the six-dimensional theory, where
 the two compact directions are mutually perpendicular.
 It is assumed that the compact radius of the fifth dimension is $R_{1}$
 and the sixth dimension is $R_{2}$.
 We begin to calculate the Casimir energy of a massless scalar field with
 periodic boundary conditions along two directions in $T^{2}$.
 The double sum of the corresponding Kaluza-Klein spectrum in 
 (\ref{eqn1}) can be expressed as
 \begin{eqnarray}
 \lefteqn{
 \sum_{m,n}{}^{\prime}
  \left(\frac{1}{R^{2}_{1}}n^{2}+\frac{1}{R^{2}_{2}}m^{2}\right)^{-s}}
 \nonumber\\
 &=&
 R^{2s}_{1}\left\{
 2\z(2s)+2\sqrt{\pi}\frac{\G\left(s-\frac{1}{2}\right)}{\G(s)}
 \left(\frac{R_{1}}{R_{2}}\right)^{1-2s}\z(2s-1)
 \right.\nonumber\\
 &&\left.\hspace{3cm}
 +\frac{8\pi^{s}}{\G(s)}\left(\frac{R_{1}}{R_{2}}\right)^{\frac{1}{2}-s}
 \sum^{\infty}_{m=1}\sum^{\infty}_{n=1}
 \left(\frac{n}{m}\right)^{s-\frac{1}{2}}
 K_{s-\frac{1}{2}}\left(2\pi \frac{R_{1}}{R_{2}}mn\right)
 \right\}
 \label{eqn2}\,,
 \end{eqnarray}
 where the prime denotes the exclusion of a zero mode.
 Here $K_{s}(x)$ is the modified Bessel function and
 $\z(x)$ is the Riemann zeta function.
 In the Appendix A the evaluation of the double sum is explicitly 
 performed. 
 Plugging into (\ref{eqn1}) leads to the Casimir energy
 \begin{eqnarray}
 V^{(P,P)}&=&-\frac{1}{64\pi^{2}R^{4}_{1}}
 \left\{
 \frac{3}{\pi^{4}}\z(5)+\frac{8\pi}{945}
                        \left(\frac{R_{1}}{R_{2}}\right)^{5}\right.
 \nonumber\\
 &&\left.\hspace{3cm}
 +\frac{16}{\pi^{2}}\left(\frac{R_{1}}{R_{2}}\right)^{\frac{5}{2}}
 \sum^{\infty}_{m=1}\sum^{\infty}_{n=1}
 \left(\frac{m}{n}\right)^{\frac{5}{2}}
 K_{-\frac{5}{2}}\left(2\pi\frac{R_{1}}{R_{2}}mn\right)
 \right\}\,,\label{eqn3}
 \end{eqnarray}
 where $(P,P)$ denotes periodic boundary conditions along two compact
 directions in $T^{2}$, the former letter corresponds to the fifth
 direction and the latter letter does the sixth direction.
 Performing the calculation of the derivative with respect to $s$
 in (\ref{eqn1}), we used the following particular values:  
 \begin{eqnarray}
 &&\z^{\prime}(-4)=\frac{3}{4\pi^{4}}\z(5)\;\;,\;\;
     \frac{\G^{\prime}(-2)}{\G(-2)^{2}}=-2\;\;,\;\;
     \G\left(-\frac{5}{2}\right)=-\frac{8}{15}\sqrt{\pi}\;\;,\nonumber\\
 &&\z(-5)=-\frac{15}{4\pi^{6}}\z(6)=-\frac{1}{252}\label{eqn4}\,.
 \end{eqnarray}
 In the Appendix B we represented the evaluation of $\z^{\prime}(-2n)$,
 where $n$ is nonnegative integer.
 By using the following form of the modified Bessel function :
 \begin{eqnarray}
 K_{-\frac{5}{2}}(z)=\sqrt{\frac{\pi}{2}}\left(z^{-\frac{1}{2}}
 +3z^{-\frac{3}{2}}+3z^{-\frac{5}{2}}
 \right)e^{-z}\,,\label{eqn5}
 \end{eqnarray}
  we can obtain 
 \begin{eqnarray}
   V^{(P,P)}&=&-\frac{1}{64\pi^{2}R^{4}_{1}}
 \left\{
 \frac{3}{\pi^{4}}\z(5)+\frac{8\pi}{945}
                        \left(\frac{R_{1}}{R_{2}}\right)^{5}
 +\frac{8L_{3}(\tau)}{\pi^{2}}\left(\frac{R_{1}}{R_{2}}\right)^{2}
 \right.\nonumber\\
 &&\left.\hspace{6cm}
 +\frac{12L_{4}(\tau)}{\pi^{3}}\frac{R_{1}}{R_{2}}
 +\frac{6L_{5}(\tau)}{\pi^{4}}
 \right\}\,,\label{eqn6}
 \end{eqnarray}
 where we defined 
 \begin{eqnarray}
 L_{3}(\tau)&\equiv& \sum^{\infty}_{m=1}m^{2}Li_{3}(q^{m})=
      \frac{1}{4}\sum^{\infty}_{n=1}
      \frac{\coth\pi\tau n}{n^{3}\sinh^{2}\pi\tau n}\label{eqn7}\,,\\
 L_{4}(\tau)&\equiv& \sum^{\infty}_{m=1}mLi_{4}(q^{m})=
      \frac{1}{4}\sum^{\infty}_{n=1}
      \frac{1}{n^{4}\sinh^{2}\pi\tau n}\label{eqn8}\,,\\
 L_{5}(\tau)&\equiv& \sum^{\infty}_{m=1}Li_{5}(q^{m})=
      \sum^{\infty}_{n=1}\frac{1}{n^{5}(e^{2\pi\tau n}-1)}\label{eqn9}
 \,,\\
 q&\equiv&e^{-2\pi\tau}\label{eqn10}\,,
 \end{eqnarray}
 where $\tau=R_{1}/R_{2}$ and
 $Li_{k}(x)=\sum^{\infty}_{n=1}x^{n}/n^{k}$ is the polylogarithm function.
 In (\ref{eqn7}), (\ref{eqn8}) and (\ref{eqn9}), the infinite sums of
 geometric sequence over $m$ are performed.

 The toroidal compactification has modular symmetry, namely,
 it is always possible to redefine the values of $R_{1}$ and $R_{2}$.
 Therefore the area of the torus ${\cal A}$ and the ratio of two radii
 $\tau$ have physical meaning.
 The ordinary modular symmetry can be specified by the three parameters,
 two radii
 $R_{1}$, $R_{2}$ and relative angle $\theta$ between two directions
 of compactification.
 In terminology of string theory, the area ${\cal A}$ corresponds to
 the K$\ddot{\rm a}$hler moduli and the modular parameter 
 $(R_{1}/R_{2})e^{i\theta}$ corresponds to the complex moduli.
 The modular invariance of the Casimir energy can be demonstrated by the
 Poisson resummation technique described in \cite{Rohm}.
 In Ref. \cite{Ponton:2001hq}, it was shown that the Casimir energy
 $V^{(P,P)}$ has extreme points at the two self-dual points
 ($\theta=\pi/2,2\pi/3$ and $R_{1}=R_{2}$) of modular symmetry.
 In the present paper it is assumed that $\theta=\pi/2$. 
 From (\ref{eqn6}), the Casimir energy can be written in terms of 
 ${\cal A}=4\pi^{2}R_{1}R_{2}$ and $\tau=R_{1}/R_{2}$ :
 \begin{eqnarray}
 V^{(P,P)}=-\frac{1}{4{\cal A}^{2}}\left\{
 \frac{8\pi^{3}}{945}\tau^{3}
 +8L_{3}(\tau)+\frac{12L_{4}(\tau)}{\pi}\frac{1}{\tau}
 +\frac{3\z(5)+6L_{5}(\tau)}{\pi^{2}}\frac{1}{\tau^{2}}
 \right\}\label{eqn11}\,.
 \end{eqnarray}
 Note that since the Casimir energy $V^{(P,P)}$ is negative for
 arbitrary $\tau$, the Casimir force due to a scalar field with periodic
 boundary conditions along two directions is attractive. 
 For instance, adopting $\tau=1\;(R_{1}=R_{2})$ to be maximal symmetry,
 the value of $V^{(P,P)}$ is approximately given by
 \begin{eqnarray}
 V^{(P,P)}\simeq -0.1502385/{\cal A}^{2}\,,\label{eqn12}
 \end{eqnarray}
 where the area is fixed.
 Thus it turns out to be attractive force.

 Next we calculate the Casimir energy of a massless scalar field
 with periodic boundary condition in the fifth direction and anti-periodic
 boundary condition in the sixth direction. 
 The double sum of the corresponding Kaluza-Klein spectrum in
 (\ref{eqn1}) can be expressed as
 \begin{eqnarray}
  \lefteqn{
 \sum_{m,n}
  \left(\frac{1}{R^{2}_{1}}n^{2}+\frac{1}{R^{2}_{2}}
  \left(m+\frac{1}{2}\right)^{2}\right)^{-s}}\nonumber\\
 &&=
 R^{2s}_{1}\left\{
 2\sqrt{\pi}\frac{\G\left(s-\frac{1}{2}\right)}{\G(s)}
 \left(\frac{R_{1}}{R_{2}}\right)^{1-2s}\z\left(2s-1,\frac{1}{2}\right)
 \right.\nonumber\\
 &&\left.
 +\frac{8\pi^{s}}{\G(s)}\left(\frac{R_{1}}{R_{2}}\right)^{\frac{1}{2}-s}
 \sum^{\infty}_{m=0}\sum^{\infty}_{n=1}
 \left(m+\frac{1}{2}\right)^{-s+\frac{1}{2}}n^{s-\frac{1}{2}}
 K_{s-\frac{1}{2}}\left(2\pi \frac{R_{1}}{R_{2}}
 \left(m+\frac{1}{2}\right)n\right)
 \right\}
 \label{eqn13}\,,
 \end{eqnarray}
 where $\z(s,\nu)$ is the Hurwitz's zeta function.
 In the Appendix A we evaluated the above double sum which corresponds
 to the case of $\alpha\rightarrow 0$ in (\ref{eqn40}).
 From (\ref{eqn1}), we can obtain
 \begin{eqnarray}
 V^{(P,A)}&=&
 -\frac{1}{64\pi^{2}R^{4}_{1}}
 \left\{
 -\frac{31\pi}{3780}
                        \left(\frac{R_{1}}{R_{2}}\right)^{5}\right.
 \nonumber\\
 &&\left.\hspace{0.2cm}
 +\frac{16}{\pi^{2}}\left(\frac{R_{1}}{R_{2}}\right)^{\frac{5}{2}}
 \sum^{\infty}_{m=0}\sum^{\infty}_{n=1}
 \left(m+\frac{1}{2}\right)^{\frac{5}{2}}n^{-\frac{5}{2}}
 K_{-\frac{5}{2}}\left(2\pi\frac{R_{1}}{R_{2}}
 \left(m+\frac{1}{2}\right)n\right)
 \right\}\,,\label{eqn14}
 \end{eqnarray}
 where we used $\z(s,1/2)=(2^{s}-1)\z(s)$ and $(P,A)$ denotes
 periodic boundary condition in the fifth direction and anti-periodic
 boundary condition in the sixth direction. 
 Using (\ref{eqn5}), the Casimir energy can be expressed as
 \begin{eqnarray}
  V^{(P,A)}=
 -\frac{1}{64\pi^{2}R^{4}_{1}}
 \left\{
 -\frac{31\pi}{3780}
                        \left(\frac{R_{1}}{R_{2}}\right)^{5}
  +\frac{8\tilde{L}_{3}(\tau)}{\pi^{2}}\left(\frac{R_{1}}{R_{2}}\right)^{2}
 +\frac{12\tilde{L}_{4}(\tau)}{\pi^{3}}\frac{R_{1}}{R_{2}}
 +\frac{6\tilde{L}_{5}(\tau)}{\pi^{4}}
 \right\}
 \,,\label{eqn15}
 \end{eqnarray}
 where we defined 
 \begin{eqnarray}
 \tilde{L}_{3}(\tau)&\equiv&
 \sum^{\infty}_{m=0}\left(m+\frac{1}{2}\right)^{2}
 Li_{3}\left(q^{m+\frac{1}{2}}\right)=\frac{1}{8}
 \sum^{\infty}_{n=1}\frac{\cosh^{2}\pi\tau n+1}
 {n^{3}\sinh^{3}\pi\tau n} \label{eqn16}\,,\\
 \tilde{L}_{4}(\tau)&\equiv& \sum^{\infty}_{m=0}\left(m+\frac{1}{2}\right)
 Li_{4}\left(q^{m+\frac{1}{2}}\right)=
 \frac{1}{4}\sum^{\infty}_{n=1}\frac{\coth\pi\tau n}
 {n^{4}\sinh\pi\tau n}\label{eqn17}\,,\\
 \tilde{L}_{5}(\tau)&\equiv& \sum^{\infty}_{m=0}
 Li_{5}\left(q^{m+\frac{1}{2}}\right)
 =\frac{1}{2}\sum^{\infty}_{n=1}\frac{1}{n^{5}\sinh\pi\tau n}
 \label{eqn18}\,.
 \end{eqnarray}
 Using the area of torus ${\cal A}$ and the ratio of two radii $\tau$,
 (\ref{eqn15}) is given by
 \begin{eqnarray}
   V^{(P,A)}=
 -\frac{1}{4{\cal A}^{2}}
 \left\{
 -\frac{31\pi^{3}}{3780}\tau^{3}
 +8\tilde{L}_{3}(\tau)
 +\frac{12\tilde{L}_{4}(\tau)}{\pi}\frac{1}{\tau}
 +\frac{6\tilde{L}_{5}(\tau)}{\pi^{2}}\frac{1}{\tau^{2}}
 \right\}
 \,.\label{eqn19}
 \end{eqnarray} 
 In the bracket of the above equation, the first term contributes to the 
 repulsive force and the remaining terms have the contributions of the
 attractive forces.
 The value of the Casimir energy for $\tau=1$, when the area is fixed,
 is approximately given by
 \begin{eqnarray}
 V^{(P,A)}\simeq +0.0139727/{\cal A}^{2}\,,\label{eqn20}
 \end{eqnarray} 
 consequently, repulsive force arises.
 In the case of $\tau\ll 1$, since the last term in (\ref{eqn19}) is
 dominant, the attractive force is generated.
 By choosing the appropriate value of $\tau$, vanishing $V^{(P,A)}$
 can be realized by balance between attractive and repulsive. 

 Successively we calculate the Casimir energy of a massless scalar field
 with anti-periodic boundary condition in the fifth direction and periodic
 boundary condition in the sixth direction. 
 By making an exchange of
 $\tau\leftrightarrow 1/\tau\;(R_{1}\leftrightarrow R_{2})$ in
 (\ref{eqn19}), we can obtain
 \begin{eqnarray}
    V^{(A,P)}=
 -\frac{1}{4{\cal A}^{2}}
 \left\{
 -\frac{31\pi^{3}}{3780}\frac{1}{\tau^{3}}
 +8\tilde{L}_{3}(\tau^{-1})
 +\frac{12\tilde{L}_{4}(\tau^{-1})}{\pi}\tau
 +\frac{6\tilde{L}_{5}(\tau^{-1})}{\pi^{2}}\tau^{2}
 \right\}\,.\label{eqn21}
 \end{eqnarray}
 As a matter of course, $V^{(P,A)}$ is equal to $V^{(A,P)}$ for $\tau=1$.
 In the case of $\tau\ll 1$, since the first term is dominant,
 it is repulsive force.

 Finally we will calculate the Casimir energy of a massless scalar field
 with anti-periodic boundary conditions in two directions.
 The double sum of the corresponding Kaluza-Klein modes in
 (\ref{eqn1}) can be given by
 \begin{eqnarray}
  \lefteqn{
 \sum_{m,n}
  \left(\frac{1}{R^{2}_{1}}\left(n+\frac{1}{2}\right)^{2}
 +\frac{1}{R^{2}_{2}}
  \left(m+\frac{1}{2}\right)^{2}\right)^{-s}}\nonumber\\
 &&=
 R^{2s}_{1}\left\{
 2\sqrt{\pi}\frac{\G\left(s-\frac{1}{2}\right)}{\G(s)}
 \left(\frac{R_{1}}{R_{2}}\right)^{1-2s}\z\left(2s-1,\frac{1}{2}\right)
 \right.\nonumber\\
 &&\left.
 +\frac{8\pi^{s}}{\G(s)}\left(\frac{R_{1}}{R_{2}}\right)^{\frac{1}{2}-s}
 \sum^{\infty}_{m=0}\sum^{\infty}_{n=1}
 \left(m+\frac{1}{2}\right)^{-s+\frac{1}{2}}n^{s-\frac{1}{2}}(-1)^{n}
 K_{s-\frac{1}{2}}\left(2\pi \frac{R_{1}}{R_{2}}
 \left(m+\frac{1}{2}\right)n\right)
 \right\}\,.\nonumber\\
 \label{eqn22}
 \end{eqnarray}
 The evaluation of the above double sum corresponds to the case of
 $\a=\b=1/2$ in the Appendix A.
 Plugging into (\ref{eqn1}), we have
 \begin{eqnarray}
   V^{(A,A)}=
 -\frac{1}{64\pi^{2}R^{4}_{1}}
 \left\{
 -\frac{31\pi}{3780}
                        \left(\frac{R_{1}}{R_{2}}\right)^{5}
  +\frac{8\hat{L}_{3}(\tau)}{\pi^{2}}\left(\frac{R_{1}}{R_{2}}\right)^{2}
 +\frac{12\hat{L}_{4}(\tau)}{\pi^{3}}\frac{R_{1}}{R_{2}}
 +\frac{6\hat{L}_{5}(\tau)}{\pi^{4}}
 \right\}
 \,,\label{eqn23}
 \end{eqnarray}
 where we defined 
 \begin{eqnarray}
 \hat{L}_{3}(\tau)&\equiv& \sum^{\infty}_{n=1}\sum^{\infty}_{m=0}
 \frac{(-1)^{n}}{n^{3}}\left(m+\frac{1}{2}\right)^{2}
 e^{-2\pi\tau n(m+1/2)}
 =-\tilde{L}_{3}(\tau)+\frac{1}{4}\tilde{L}_{3}(2\tau)\,,\label{eqn24}\\
 \hat{L}_{4}(\tau)&\equiv&
  \sum^{\infty}_{n=1}\sum^{\infty}_{m=0}
 \frac{(-1)^{n}}{n^{4}}\left(m+\frac{1}{2}\right)
 e^{-2\pi\tau n(m+1/2)}
 =-\tilde{L}_{4}(\tau)+\frac{1}{8}\tilde{L}_{4}(2\tau)\,,\label{eqn25} \\
 \hat{L}_{5}(\tau)&\equiv& 
 \sum^{\infty}_{n=1}\sum^{\infty}_{m=0}
 \frac{(-1)^{n}}{n^{5}}
 e^{-2\pi\tau n(m+1/2)}
 =-\tilde{L}_{5}(\tau)+\frac{1}{16}\tilde{L}_{5}(2\tau)\,.\label{eqn26}
 \end{eqnarray}
 Here we rewrote the above sums by using (\ref{eqn16}), (\ref{eqn17})
 and (\ref{eqn18}).
 The Casimir energy with anti-periodic boundary conditions in two
 directions can be rewritten in terms of ${\cal A}$ and $\tau$ as follows
 \begin{eqnarray}
 V^{(A,A)}= -\frac{1}{4{\cal A}^{2}}
 \left\{
 -\frac{31\pi^{3}}{3780}\tau^{3}
 +8\hat{L}_{3}(\tau)
 +\frac{12\hat{L}_{4}(\tau)}{\pi}\frac{1}{\tau}
 +\frac{6\hat{L}_{5}(\tau)}{\pi^{2}}\frac{1}{\tau^{2}}
 \right\}\label{eqn27}\,.
 \end{eqnarray}
 Since $\tilde{L}_{3}$, $\tilde{L}_{4}$ and $\tilde{L}_{5}$
 are decreasing for $\tau$,
 $\hat{L}_{3}$, $\hat{L}_{4}$ and $\hat{L}_{5}$ are always
 negative.
 Namely $V^{(A,A)}$ is positive for arbitrary $\tau$.
 Therefore it generates repulsive force.
 For instance, in the case of $\tau=1$ when ${\cal A}$ is fixed, the
 value of $V^{(A,A)}$ is approximately
 \begin{eqnarray}
 V^{(A,A)}\simeq +0.1126133/{\cal A}^{2}\,,\label{eqn28}
 \end{eqnarray}
 thus it is repulsive force.

 Consequently, $V^{(P,P)}$ and $V^{(A,A)}$ can generate the attractive
 force and repulsive force, respectively.
 Whether $V^{(P,A)}$ and $V^{(A,P)}$ generate attractive
 or repulsive forces depends on the value of $\tau$.
 We calculated the Casimir energies of a single massless scalar field in
 the present paper, while for fermion field we need to multiply by $-1$.
 Taking into account of the number of degrees of freedom in matter fields
 with various boundary conditions, the contributions of the matter fields
 are given by the sum of
 (\ref{eqn11}), (\ref{eqn19}), (\ref{eqn21}) and (\ref{eqn27}).
 Thus the total Casimir energy due to the matter fields depends on
 model-building.
 Moreover we need to add the contribution of a bulk cosmological
 constant :
 $\dis\int d^{6}x\sqrt{G}\;\L=\int d^{4}x\;\L{\cal A}$,
 where $\L$ is a bulk cosmological constant which corresponds to the
 vacuum energy.
 Accordingly it is possible for the total Casimir energy to have
 minimum for ${\cal A}$ and $\tau$.
 However, the stability of the system depends on the
 sign of $\L$ which leads to the familiar cosmological constant problem.
 In the present paper we do not mention the problem.
 Setting to a certain supersymmetric model,
 the preservation of supersymmetry via toroidal compactification
 guarantees vanishing Casimir energy \cite{Ponton:2001hq}.

%
 \section{Application to $Z^{2}$ orbifold}

 We shall discuss a procedure of the application to $T^{2}/Z_{2}$
 orbifold.
 The matter fields with boundary conditions of
 $(P,P)$, $(P,A)$, $(A,P)$ and $(A,A)$ can be decomposed by $Z_{2}$
 orbifold.
 The wave function of each matter field can be separated into the cosine
 function and the sine function by performing the operation of the orbifold.
 Equivalently, this implies $Z_{2}$ projections of the following
 three transformations
 \begin{eqnarray}
 r_{1}&:&(x_{5},x_{6})\mapsto (x_{5}+2\pi R_{1},x_{6})\,,
 \label{eqn29}\\
 r_{2}&:&(x_{5},x_{6})\mapsto (x_{5},x_{6}+2\pi R_{2})\,,\label{eqn30}\\
 r_{3}&:&(x_{5},x_{6})\mapsto (-x_{5},-x_{6})\,,\label{eqn31}
 \end{eqnarray}
 where $x_{5}$ and $x_{6}$ denotes the fifth coordinate and the sixth
 coordinate, respectively.
 The $Z_{2}$ parity of matter fields via these transformations can be
 specified by $(r_{1},r_{2},r_{3})$, where
 $r_{1},r_{2},r_{3}=\pm 1$ under the orbifold symmetry.
 Namely this means that $+1$ corresponds to the even state and
 $-1$ odd state.
 The illustration of the decompositions via $Z_{2}$ projection can be
 represented as follows
 \begin{eqnarray}
 (P,P)&:&{\rm exp}\left(i\frac{x_{5}}{R_{1}}m+i\frac{x_{6}}{R_{2}}n\right)
 \rightarrow
 \left\{ \begin{array}{l}
 \dis\cos\left(\frac{x_{5}}{R_{1}}m+\frac{x_{6}}{R_{2}}n\right)\\
 \dis\sin\left(\frac{x_{5}}{R_{1}}m+\frac{x_{6}}{R_{2}}n\right)
 \end{array} \right.\,,\label{eqn32}\\
 (P,A)&:&{\rm exp}\left(i\frac{x_{5}}{R_{1}}m+i\frac{x_{6}}{R_{2}}
 \left(n+\frac{1}{2}\right)\right)
 \rightarrow
 \left\{ \begin{array}{l}
 \dis\cos\left(\frac{x_{5}}{R_{1}}m
               +\frac{x_{6}}{R_{2}}\left(n+\frac{1}{2}\right)\right)\\
 \dis\sin\left(\frac{x_{5}}{R_{1}}m
               +\frac{x_{6}}{R_{2}}\left(n+\frac{1}{2}\right)\right)
 \end{array} \right.\,,\label{eqn33}\\
 (A,P)&:&{\rm exp}\left(i\frac{x_{5}}{R_{1}}\left(m+\frac{1}{2}\right)
 +i\frac{x_{6}}{R_{2}}n\right)
 \rightarrow
 \left\{ \begin{array}{l}
 \dis\cos\left(\frac{x_{5}}{R_{1}}\left(m+\frac{1}{2}\right)
               +\frac{x_{6}}{R_{2}}n\right)\\
 \dis\sin\left(\frac{x_{5}}{R_{1}}\left(m+\frac{1}{2}\right)
               +\frac{x_{6}}{R_{2}}n\right)
 \end{array} \right.\,,\label{eqn34}\\
 (A,A)&:&{\rm exp}\left(i\frac{x_{5}}{R_{1}}\left(m+\frac{1}{2}\right)
 +i\frac{x_{6}}{R_{2}}\left(n+\frac{1}{2}\right)\right)\nonumber\\
 &&\hspace{4cm}\rightarrow
 \left\{ \begin{array}{l}
 \dis\cos\left(\frac{x_{5}}{R_{1}}\left(m+\frac{1}{2}\right)
               +\frac{x_{6}}{R_{2}}\left(n+\frac{1}{2}\right)\right)\\
 \dis\sin\left(\frac{x_{5}}{R_{1}}\left(m+\frac{1}{2}\right)
               +\frac{x_{6}}{R_{2}}\left(n+\frac{1}{2}\right)\right)
 \end{array} \right.\,,\label{eqn35}
 \end{eqnarray}
 where the symbol $\rightarrow$ denotes the operation of $Z_{2}$
 orbifold.
 Here we omitted the normalization factors and the four-dimensional parts
 in matter fields.
 Under the orbifold symmetry, note that $(+1,+1,+1)$ and $(+1,+1,-1)$
 states result from $(P,P)$, $(+1,-1,+1)$ and $(+1,-1,-1)$
 from $(P,A)$, $(-1,+1,+1)$ and $(-1,+1,-1)$
 from $(A,P)$, $(-1,-1,+1)$ and $(-1,-1,-1)$ from $(A,A)$.    
 Thus the $Z_{2}$ orbifold can produce the eight states.
 Furthermore there are four orbifold fixed planes on $T^{2}/Z_{2}$ at
 $(x_{5},x_{6})=(0,0),(\pi R_{1},0),(0,\pi R_{2}),(\pi R_{1},\pi R_{2})$.
 There exist extra contributions of
 the localized terms on these fixed planes, for example, kinetic terms, 
 mass terms and interaction terms between bulk and brane as well as
 brane tension. 
 Since the Kaluza-Klein spectrum is modified by these brane-localized
 terms, these effects must be considered.
 For example, in five-dimensional $S^{1}/Z_{2}$ model,
 the Casimir energy including brane-localized kinetic terms had been
 calculated \cite{Ponton:2001hq}.
 Consequently the total Casimir energy consists of the matter fields, a bulk
 cosmological constant as well as brane-localized terms.
 It is considered that the Casimir energy including all contributions is
 very complicated form, and minimum problem of the Casimir energy is
 closely related to the cosmological constant problem.
 We are going to investigate the points elsewhere.

 Adopting a certain model on $T^{2}/Z_{2}$ orbifold, the Casimir
 energies due to the matter fields will be calculated by taking account
 of the
 number of the degrees of freedom and $(r_{1},r_{2},r_{3})$ parity
 assignments given in matter content of the model when neglecting the
 effects of branes.
 When performing the calculation,
 we need to make the replacements $R_{1}\rightarrow R_{1}/2$ and
 $R_{2}\rightarrow R_{2}/2$, simultaneously, multiply $1/2\times 1/2$
 factor via half modes over $m,n$.
 Therefore we can obtain the following forms
 \begin{eqnarray}
 V^{(++\pm)}&=& \frac{1}{4}\;V^{(P,P)}\left(\frac{R_{1}}{2},\frac{R_{2}}{2}
                \right)\nonumber\\
 V^{(+-\pm)}&=& \frac{1}{4}\;V^{(P,A)}\left(\frac{R_{1}}{2},\frac{R_{2}}{2}
                \right) \nonumber\\
 V^{(-+\pm)}&=& \frac{1}{4}\;V^{(A,P)}\left(\frac{R_{1}}{2},\frac{R_{2}}{2}
                \right)\nonumber\\
 V^{(--\pm)}&=& \frac{1}{4}\;V^{(A,A)}\left(\frac{R_{1}}{2},\frac{R_{2}}{2}
                \right)
 \end{eqnarray}
 When considering the radius stabilization of the concrete model on
 $T^{2}/Z_{2}$ orbifold, it is important to use results obtained here.
 The concrete model is beyond the scope of this paper and we do not
 mention it here.

 \section{Summary and Discussion}

 We calculated the Casimir energies due to a massless scalar field with
 various boundary conditions on $T^{2}$ compactification, assuming that
 two compact directions are perpendicular each other.
 The Casimir energies can be explicitly represented in terms of the area
 of torus and the ratio of two radii. 
 Consequently, it was shown that the case of $(P,P)$ boundary condition
 is attractive force and the case of $(A,A)$ boundary condition is
 repulsive force.
 For $(P,A)$ and $(A,P)$, whether attractive or repulsive depends on
 the ratio of two radii.
 For fermion field, opposite force works.

 When calculating the Casimir energies of the matter fields on 
 a certain $T^{2}/Z_{2}$ orbifold model,
 we need to multiply our results to the number of degrees of freedom 
 assigning eight kinds of $Z_{2}$ parity $(\pm 1,\pm 1,\pm 1)$ of matter
 content.  
 Thus the contributions of the matter fields in radius stabilization of the
 model will be explicitly evaluated.
 Furthermore there are contributions of a bulk cosmological constant and
 brane-localized terms (kinetic term, mass term, interaction terms and
 brane tension) on the orbifold fixed planes.
 Since Kaluza-Klein spectrum is modified by these brane-localized terms,
 it can be considered that the total Casimir energy is very 
 complicated form.
 That the total Casimir energy has a minimum point for the area of the
 torus and the ratio of two radii is related to the cosmological
 constant problem.
 We are going to describe it elsewhere.
 In the present paper we could provide the formulas of the Casimir
 energies with various boundary conditions when considering the
 radius stabilization in the model.

 \section*{Appendix A: Evaluation of double sums}

 Calculating the Casimir energy for field with various boundary
 conditions on toroidal compactification,
 we encounter the double summation of infinite series \cite{zeta1}.

 When we compute the Casimir energy of a massless scalar field with
 periodic boundary conditions along two compact directions in $T^{2}$,
 we must evaluate the following double sum
 \begin{eqnarray}
 I(a;s)=
 \sum_{m,n}{}^{\prime}\left[n^{2}+a^{2}m^{2}\right]^{-s}\label{eqn36}\,,
 \end{eqnarray}
 where the prime denotes $(m,n)\neq(0,0)$.
 The double sum can be decomposed as follows
 \begin{eqnarray}
 \sum_{n}{}^{\prime}n^{-2s}
 +\sum_{m}{}^{\prime}\sum_{n}\left[n^{2}+a^{2}m^{2}\right]^{-s}\,.
 \label{eqn37}
 \end{eqnarray}
 The first term can be written in terms of the Riemann zeta function.
 The second term can be rewritten by using the gamma function what
 is called Mellin transformation.
 Consequently, we obtain \cite{Ponton:2001hq}
 \begin{eqnarray}
 I(a;s)
 &=&2\z(2s)+\sum_{m}{}^{\prime}\sum_{n}\frac{1}{\G(s)}
    \int^{\infty}_{0}dt\;t^{s-1}e^{-\left(n^{2}+a^{2}m^{2}\right)t}
 \nonumber\\
 &=&2\z(2s)+\sum_{m}{}^{\prime}\frac{\sqrt{\pi}}{\G(s)}
     \int^{\infty}_{0}dt\;t^{s-\frac{3}{2}}e^{-a^{2}m^{2}t}
     \left(1+2\sum^{\infty}_{n=1}e^{-\frac{\pi^{2}}{t}n^{2}}\right)
 \nonumber\\
 &=&2\z(2s)+2\sqrt{\pi}\frac{\G\left(s-\frac{1}{2}\right)}{\G(s)}
    |a|^{1-2s}\z(2s-1)\nonumber\\
 &&\hspace{2cm}
    +\frac{8\pi^{s}}{\G(s)}|a|^{\frac{1}{2}-s}
    \sum^{\infty}_{m=1}\sum^{\infty}_{n=1}
    \left(\frac{n}{m}\right)^{s-\frac{1}{2}}
    K_{s-\frac{1}{2}}\left(2\pi |a|mn\right)\,,\label{eqn38}
 \end{eqnarray}
 where $K_{s}(x)$ is the modified Bessel function, we used the Poisson 
 re-summation formula in the second line.

 Next we shall evaluate the double sum with non-periodic boundary
 conditions as follows
 \begin{eqnarray}
 I(a;\a,\b;s)=
 \sum_{m,n}\left[(n+\a)^{2}+a^{2}(m+\b)^{2}\right]^{-s}\,,\label{eqn39}
 \end{eqnarray}
 where $0< \a,\b<1$.
 Following the similar procedures in (\ref{eqn38}), we obtain 
 \begin{eqnarray}
 \lefteqn{I(a;\a,\b;s)}\nonumber\\
 &=&
 \sum_{m,n}\frac{1}{\G(s)}\int^{\infty}_{0}dt\;
 t^{s-1}e^{-\left((n+\a)^{2}+a^{2}(m+\b)^{2}\right)t}\nonumber\\
 &=&
 \frac{\sqrt{\pi}}{\G(s)}\sum^{\infty}_{m=-\infty}
 \int^{\infty}_{0}t^{s-\frac{3}{2}}e^{-a^{2}\left(m+\b\right)^{2}t}
 \left(1+2\sum^{\infty}_{n=1}\cos\left(2\pi n\a\right)
 e^{-\frac{\pi^{2}}{t}n^{2}}\right)\nonumber\\
 &=&
 \sqrt{\pi}\frac{\G\left(s-\frac{1}{2}\right)}{\G(s)}
 |a|^{1-2s}\left(\frac{}{}\z(2s-1,\b)+\z(2s-1,1-\b)\right)
 \nonumber\\
 &&+\frac{4\pi^{s}}{\G(s)}|a|^{\frac{1}{2}-s}
 \sum^{\infty}_{m=0}\sum^{\infty}_{n=1}
 n^{s-\frac{1}{2}}\cos\left(2\pi n\a\right)
 \left\{\frac{}{}
 (m+\b)^{-s+\frac{1}{2}}K_{s-\frac{1}{2}}\left(2\pi |a|(m+\b)n\right)
 \right.\nonumber\\
 &&\left.\hspace{4cm}
 +(m+1-\b)^{-s+\frac{1}{2}}K_{s-\frac{1}{2}}\left(2\pi |a|(m+1-\b)n\right)
 \frac{}{}\right\}\,,\label{eqn40}
 \end{eqnarray}
 where
 \begin{eqnarray}
 \z(s,\nu)=\sum^{\infty}_{n=0}\frac{1}{(n+\nu)^{s}}\label{eqn41}
 \end{eqnarray}
 is the Hurwitz's zeta function.

 \section*{Appendix B: Evaluation of $\zeta^{\prime}(-2n)$}

 We derive the relation between $\zeta^{\prime}(-2n)$ and $\zeta(2n+1)$
 by calculating the Casimir energy for ${\rm R}^{2n}\times S^{1}$,
 where $n$ is nonnegative integer.
 We consider a massless scalar field with periodic boundary condition
 on the $S^{1}$ compactification with radius $L$.
 The Casimir energy $E$ is given by
 \begin{eqnarray}
 E&=&\frac{1}{2}\sum^{\infty}_{m=-\infty}{}^{\prime}
     \int\frac{d^{2n}k}{(2\pi)^{2n}}
   \log\left(k^{2}+\frac{m^{2}}{L^{2}}\right)\nonumber\\
 &=&-\frac{1}{2}\left.\frac{\p}{\p s}\right|_{s=0}
     \sum^{\infty}_{m=-\infty}{}^{\prime}\int\frac{d^{2n}k}{(2\pi)^{2n}}
     \left(k^{2}+\frac{m^{2}}{L^{2}}\right)^{-s}\nonumber\\
 &=&-\frac{1}{2}\left.\frac{\p}{\p s}\right|_{s=0}
     \sum^{\infty}_{m=-\infty}{}^{\prime}\int\frac{d^{2n}k}{(2\pi)^{2n}}
    \frac{1}{\G(s)}\int^{\infty}_{0}dt\;
    e^{-\left(k^{2}+\frac{m^{2}}{L^{2}}\right)t}t^{s-1}\nonumber\\
 &=&-\frac{\pi^{n}}{(2\pi)^{2n}L^{2n}}
    \left.\frac{\p}{\p s}\right|_{s=0}
    \frac{\G(s-n)}{\G(s)}L^{2s}\z(2s-2n)\nonumber\\
 &=&
 -\frac{2(-1)^{n}\pi^{n}}{(2\pi)^{2n}\G(n+1)L^{2n}}\z^{\prime}(-2n)
 \,.\label{eqn42}
 \end{eqnarray}

 Next we can rewrite the first line of (\ref{eqn42}) as follows
 \begin{eqnarray}
 E&=&-\frac{1}{2}
     \sum^{\infty}_{m=-\infty}{}^{\prime}\int\frac{d^{2n}k}{(2\pi)^{2n}}
     \int^{\infty}_{0}ds\;\frac{1}{s}
     e^{-\left(k^{2}+\frac{m^{2}}{L^{2}}\right)s}\nonumber\\
 &=&-\frac{\pi^{n}}{2(2\pi)^{2n}}\int^{\infty}_{0}ds\;\frac{1}{s^{n+1}}
     \sum^{\infty}_{m=-\infty}{}^{\prime}e^{-\frac{m^{2}}{L^{2}}s}
     \nonumber\\
 &=&-\frac{\pi^{n}}{2(2\pi)^{2n}}\int^{\infty}_{0}ds\;\frac{1}{s^{n+1}}
     \sqrt{\frac{\pi L^{2}}{s}}
     \sum^{\infty}_{m=-\infty}{}^{\prime}e^{-\frac{\pi^{2}L^{2}}{s}m^{2}}
     \nonumber\\
 &=&-\frac{1}{2^{2n}\pi^{3n+\frac{1}{2}}L^{2n}}\G\left(n+\frac{1}{2}\right)
     \z(2n+1)\label{eqn43}
 \end{eqnarray}
 where we used the Poisson re-summation formula in the third line.
 Since (\ref{eqn42}) is equal to (\ref{eqn43}), we obtain
 \begin{eqnarray}
 \z^{\prime}(-2n)
 &=&\frac{(-1)^{n}}{2\pi^{2n+\frac{1}{2}}}\G(n+1)
   \G\left(n+\frac{1}{2}\right)
     \z(2n+1)\nonumber\\
 &=&\frac{(-1)^{n}}{2(2\pi)^{2n}}\G(2n+1)\z(2n+1)\,.\label{eqn44}
 \end{eqnarray}
 Below we tabulate the particular values of $\z^{\prime}(-2n)$,
 for instance, for $n=1,2,3,4$.
 \begin{eqnarray}
 \z^{\prime}(-2)&=& -\frac{1}{4\pi^{2}}\z(3)\sim -0.03048\nonumber\,,\\
 \z^{\prime}(-4)&=& \frac{3}{4\pi^{4}}\z(5)\sim 0.0080\nonumber\,,\\
 \z^{\prime}(-6)&=& -\frac{45}{8\pi^{6}}\z(7)\sim -0.00592\nonumber\,,\\
 \z^{\prime}(-8)&=& \frac{315}{4\pi^{8}}\z(9)\sim 0.00835\,.\label{eqn45}
 \end{eqnarray}

 Next, for your information, we can obtain a specific value of $n=0$ by
 using the following identity \cite{whittaker}
 \begin{eqnarray}
 \z(1-z)=2^{1-z}\pi^{-z}\cos\frac{z\pi}{2}\G(z)\z(z)
 \,.\label{eqn46}
 \end{eqnarray}
 Performing the logarithmic derivative with respect to $z$, we get
 \begin{eqnarray}
 \frac{\z^{\prime}(1-z)}{\z(1-z)}=\log(2\pi)
  +\frac{\pi}{2}\tan\frac{z\pi}{2}-\psi(z)-\frac{\z^{\prime}(z)}{\z(z)}
 \,,\label{eqn47}
 \end{eqnarray}
 where $\psi(z)$ is the polygamma function.
 Taking the limit of $z\rightarrow 1$, we have
 \begin{eqnarray}
 \frac{\z^{\prime}(0)}{\z(0)}=\log(2\pi)-\psi(1)
 +\lim_{z\rightarrow 1}
  \left(-\frac{\z^{\prime}(z)}{\z(z)}+\frac{\pi}{2}\tan\frac{z\pi}{2}
 \right)=\log(2\pi)\,.\label{eqn48}
 \end{eqnarray}
 Here we used $\psi(1)=-\gamma$ and
 $\z(z)=1/(z-1)+\gamma+{\cal O}(|z-1|)$ for $z\sim 1$, where $\gamma$
 is the Euler constant.
 Therefore we have
 \begin{eqnarray}
 \z^{\prime}(0)=\z(0)\log(2\pi)=-\frac{1}{2}\log(2\pi)\,.\label{eqn49}
 \end{eqnarray}
%
%


\begin{thebibliography}{99}
%
 \bibitem{Kaluza}
 Th. Kaluza, Situngsber. Preuss. Akad. Wiss. Phys. Math. K1(1921)966
%
 \bibitem{Klein}
 O. Klein, Nature 118(1926)516, Z. Phys. 37(1926)895
%
  \bibitem{KK}
 T. Appelquist, A. Chodos and P. G. Freund,
 ``Modern Kaluza-Klein Theories'',
 {\it Addison-Wessley, USA (1987)} 
%
 \bibitem{Appelquist:1983vs}
 T.~Appelquist and A.~Chodos,
 ``The Quantum Dynamics Of Kaluza-Klein Theories,''
 Phys.\ Rev.\ D {\bf 28}, 772 (1983).
%
 \bibitem{Antoniadis:1998ig}
 I.~Antoniadis, N.~Arkani-Hamed, S.~Dimopoulos and G.~R.~Dvali,
 ``New dimensions at a millimeter to a Fermi and superstrings at a TeV,''
 Phys.\ Lett.\ B {\bf 436}, 257 (1998)
[arXiv:hep-ph/9804398].
%
 \bibitem{Antoniadis:1990ew}
 I.~Antoniadis,
 ``A Possible New Dimension At A Few Tev,''
 Phys.\ Lett.\ B {\bf 246}, 377 (1990).
%
 \bibitem{Arkani-Hamed:1998rs}
 N.~Arkani-Hamed, S.~Dimopoulos and G.~R.~Dvali,
 ``The hierarchy problem and new dimensions at a millimeter,''
 Phys.\ Lett.\ B {\bf 429}, 263 (1998)
 [arXiv:hep-ph/9803315].
%
 \bibitem{Randall:1999ee}
 L.~Randall and R.~Sundrum,
 ``A large mass hierarchy from a small extra dimension,''
 Phys.\ Rev.\ Lett.\  {\bf 83}, 3370 (1999) [hep-ph/9905221].
%
 \bibitem{Randall:1999vf}
 L.~Randall and R.~Sundrum,
 ``An alternative to compactification,''
 Phys.\ Rev.\ Lett.\  {\bf 83}, 4690 (1999) [hep-th/9906064].
%
 \bibitem{Kawamura:1999nj}
 Y.~Kawamura,
 ``Gauge symmetry reduction from the extra space S(1)/Z(2),''
 Prog.\ Theor.\ Phys.\  {\bf 103}, 613 (2000)
 [arXiv:hep-ph/9902423].
%
 \bibitem{Barbieri:2000vh}
 R.~Barbieri, L.~J.~Hall and Y.~Nomura,
 ``A constrained standard model from a compact extra dimension,''
 Phys.\ Rev.\ D {\bf 63}, 105007 (2001)
 [arXiv:hep-ph/0011311].
%
 \bibitem{Hofmann:2000cj}
 R.~Hofmann, P.~Kanti and M.~Pospelov,
 ``(De-)stabilization of an extra dimension due to a Casimir force,''
 Phys.\ Rev.\ D {\bf 63}, 124020 (2001)
 [arXiv:hep-ph/0012213].
%
 \bibitem{Ponton:2001hq}
 E.~Ponton and E.~Poppitz,
 ``Casimir energy and radius stabilization in five and six dimensional
 orbifolds,''JHEP {\bf 0106}, 019 (2001)
 [arXiv:hep-ph/0105021].
%
 \bibitem{zeta1}
 E. Elizalde, S. D. Odintsov, A. Romeo, A. A. Bytsenko and S. Zerbini,
 ``Zeta Regularization Technique with Applications'', 
 {\it World Scientific, Singapore (1994)}.
%
 \bibitem{zeta2}
 S. W. Hawking, Commun. Math. Phys, {\bf 55}, 133 (1977).
%
 \bibitem{Elizalde:fg}
 E.~Elizalde,
 ``Multiple Zeta Functions With Arbitrary Exponents,''
 J.\ Phys.\ A {\bf 22}, 931 (1989).
%
 \bibitem{Itzykson:xw}
 C.~Itzykson and J.~M.~Luck,
 ``Arithmetical Degeneracies In Simple Quantum Systems,''
 J.\ Phys.\ A {\bf 19}, 211 (1986).
%
 \bibitem{kubota}
 T. Kubota, ``Elementary theory of Eisenstein Series'',
 {\it Kodanha and Halsted Press (1973)}
%
 \bibitem{Rohm}
 R. Rohm,
 ``Spontaneous supersymmetry breaking in supersymmetric string
 theories,''Nucl.\ Phys.\ B {\bf 237}, 553 (1984).
%
  \bibitem{whittaker}
 E. T. Whittaker and G. N. Watson, ``Course of Modern Analysis'',
 {\it Cambridge University Press (1996)}
%
\end{thebibliography}
\end{document}